\documentclass[10pt,conference,letterpaper,nofonttune]{IEEEtran}
\usepackage{url}
\usepackage[utf8]{inputenc}
\usepackage{amsmath}
\usepackage{amssymb}
\usepackage{xspace}
\usepackage{epsfig}
\usepackage{balance}
\usepackage{listings}
\usepackage[dvipsnames]{xcolor}

\definecolor{codegray}{rgb}{0.25,0.25,0.25}
\definecolor{codepurple}{rgb}{0.58,0,0.82}
\lstdefinestyle{mystyle}{
  commentstyle=\color{PineGreen},
  keywordstyle=\color{MidnightBlue},
  numberstyle=\tiny\color{codegray},
  stringstyle=\color{codepurple},
  basicstyle=\ttfamily\footnotesize,
  breakatwhitespace=true,         
  breaklines=true,                 
  captionpos=b,
  frame=tb,
  keepspaces=true,                 
  numbers=left,                    
  numbersep=5pt,                  
  showspaces=false,                
  showstringspaces=false,
  showtabs=false,                  
  tabsize=2,
  xleftmargin=10pt,
  belowskip=-10pt,
  float=htbp,  
}
 
\lstdefinelanguage{mybash}{%
  language     = bash,
  morekeywords = {docker,python3},
}

\usepackage[acronyms,nonumberlist,nopostdot,nomain,nogroupskip,acronymlists={hidden}]{glossaries}
\newglossary[algh]{hidden}{acrh}{acnh}{Hidden Acronyms}

\usepackage{booktabs}
\usepackage{tabularx}

\usepackage{tikz}
\usepackage{pgfplots}
\pgfplotsset{compat=newest}
\pgfplotsset{plot coordinates/math parser=false}
\newlength\fheight
\newlength\fwidth
\usetikzlibrary{plotmarks,patterns,decorations.pathreplacing,backgrounds,calc,arrows,arrows.meta,spy,matrix,scopes}
\usepgfplotslibrary{patchplots,groupplots}
\usepackage{tikzscale}
\usepackage{hyperref}

\newif\ifexttikz
\exttikztrue

\ifexttikz
  \usetikzlibrary{external}
\fi

\usepackage{multirow}

\usepackage[font=footnotesize]{subcaption}
\usepackage[font=footnotesize]{caption}

\usepackage{mathtools}
\usepackage[numbers,sort&compress]{natbib}

\usepackage{soul}

\newacronym{3gpp}{3GPP}{3rd Generation Partnership Project}
\newacronym{4g}{4G}{4th generation}
\newacronym{5g}{5G}{5th generation}
\newacronym{6g}{6G}{6th generation}
\newacronym{5gc}{5GC}{5G Core}
\newacronym{adc}{ADC}{Analog to Digital Converter}
\newacronym{aerpaw}{AERPAW}{Aerial Experimentation and Research Platform for Advanced Wireless}
\newacronym{ai}{AI}{Artificial Intelligence}
\newacronym{aimd}{AIMD}{Additive Increase Multiplicative Decrease}
\newacronym{am}{AM}{Acknowledged Mode}
\newacronym{amc}{AMC}{Adaptive Modulation and Coding}
\newacronym{amf}{AMF}{Access and Mobility Management Function}
\newacronym{aops}{AOPS}{Adaptive Order Prediction Scheduling}
\newacronym{api}{API}{Application Programming Interface}
\newacronym{apn}{APN}{Access Point Name}
\newacronym{ap}{AP}{Application Protocol}
\newacronym{aqm}{AQM}{Active Queue Management}
\newacronym{ausf}{AUSF}{Authentication Server Function}
\newacronym{avc}{AVC}{Advanced Video Coding}
\newacronym{awgn}{AGWN}{Additive White Gaussian Noise}
\newacronym{balia}{BALIA}{Balanced Link Adaptation Algorithm}
\newacronym{bbu}{BBU}{Base Band Unit}
\newacronym{bdp}{BDP}{Bandwidth-Delay Product}
\newacronym{ber}{BER}{Bit Error Rate}
\newacronym{bf}{BF}{Beamforming}
\newacronym{bler}{BLER}{Block Error Rate}
\newacronym{brr}{BRR}{Bayesian Ridge Regressor}
\newacronym{bs}{BS}{Base Station}
\newacronym{bsr}{BSR}{Buffer Status Report}
\newacronym{bss}{BSS}{Business Support System}
\newacronym{ca}{CA}{Carrier Aggregation}
\newacronym{caas}{CaaS}{Connectivity-as-a-Service}
\newacronym{cb}{CB}{Code Block}
\newacronym{cc}{CC}{Congestion Control}
\newacronym{ccid}{CCID}{Congestion Control ID}
\newacronym{cco}{CC}{Carrier Component}
\newacronym{cdd}{CDD}{Cyclic Delay Diversity}
\newacronym{cdf}{CDF}{Cumulative Distribution Function}
\newacronym{cdn}{CDN}{Content Distribution Network}
\newacronym{cn}{CN}{Core Network}
\newacronym{codel}{CoDel}{Controlled Delay Management}
\newacronym{comac}{COMAC}{Converged Multi-Access and Core}
\newacronym{cord}{CORD}{Central Office Re-architected as a Datacenter}
\newacronym{cornet}{CORNET}{COgnitive Radio NETwork}
\newacronym{cosmos}{COSMOS}{Cloud Enhanced Open Software Defined Mobile Wireless Testbed for City-Scale Deployment}
\newacronym{cots}{COTS}{Commercial Off-the-Shelf}
\newacronym{cp}{CP}{Control Plane}
\newacronym{cyp}{CP}{Cyclic Prefix}
\newacronym{up}{UP}{User Plane}
\newacronym{cpu}{CPU}{Central Processing Unit}
\newacronym{cqi}{CQI}{Channel Quality Information}
\newacronym{cr}{CR}{Cognitive Radio}
\newacronym{cran}{CRAN}{Cloud \gls{ran}}
\newacronym{crs}{CRS}{Cell Reference Signal}
\newacronym{csi}{CSI}{Channel State Information}
\newacronym{csirs}{CSI-RS}{Channel State Information - Reference Signal}
\newacronym{cu}{CU}{Central Unit}
\newacronym{d2tcp}{D$^2$TCP}{Deadline-aware Data center TCP}
\newacronym{d3}{D$^3$}{Deadline-Driven Delivery}
\newacronym{dac}{DAC}{Digital to Analog Converter}
\newacronym{dag}{DAG}{Directed Acyclic Graph}
\newacronym{das}{DAS}{Distributed Antenna System}
\newacronym{dash}{DASH}{Dynamic Adaptive Streaming over HTTP}
\newacronym{dc}{DC}{Dual Connectivity}
\newacronym{dccp}{DCCP}{Datagram Congestion Control Protocol}
\newacronym{dce}{DCE}{Direct Code Execution}
\newacronym{dci}{DCI}{Downlink Control Information}
\newacronym{dctcp}{DCTCP}{Data Center TCP}
\newacronym{dl}{DL}{Downlink}
\newacronym{dmr}{DMR}{Deadline Miss Ratio}
\newacronym{dmrs}{DMRS}{DeModulation Reference Signal}
\newacronym{drlcc}{DRL-CC}{Deep Reinforcement Learning Congestion Control}
\newacronym{drs}{DRS}{Discovery Reference Signal}
\newacronym{du}{DU}{Distributed Unit}
\newacronym{e2e}{E2E}{end-to-end}
\newacronym{ecaas}{ECaaS}{Edge-Cloud-as-a-Service}
\newacronym{ecn}{ECN}{Explicit Congestion Notification}
\newacronym{edf}{EDF}{Earliest Deadline First}
\newacronym{embb}{eMBB}{Enhanced Mobile Broadband}
\newacronym{empower}{EMPOWER}{EMpowering transatlantic PlatfOrms for advanced WirEless Research}
\newacronym{enb}{eNB}{evolved Node Base}
\newacronym{endc}{EN-DC}{E-UTRAN-\gls{nr} \gls{dc}}
\newacronym{epc}{EPC}{Evolved Packet Core}
\newacronym{eps}{EPS}{Evolved Packet System}
\newacronym{es}{ES}{Edge Server}
\newacronym{etsi}{ETSI}{European Telecommunications Standards Institute}
\newacronym[firstplural=Estimated Times of Arrival (ETAs)]{eta}{ETA}{Estimated Time of Arrival}
\newacronym{eutran}{E-UTRAN}{Evolved Universal Terrestrial Access Network}
\newacronym{faas}{FaaS}{Function-as-a-Service}
\newacronym{fapi}{FAPI}{Functional Application Platform Interface}
\newacronym{fdd}{FDD}{Frequency Division Duplexing}
\newacronym{fdm}{FDM}{Frequency Division Multiplexing}
\newacronym{fdma}{FDMA}{Frequency Division Multiple Access}
\newacronym{fed4fire}{FED4FIRE+}{Federation 4 Future Internet Research and Experimentation Plus}
\newacronym{fir}{FIR}{Finite Impulse Response}
\newacronym{fit}{FIT}{Future \acrlong{iot}}
\newacronym{fpga}{FPGA}{Field Programmable Gate Array}
\newacronym{fr2}{FR2}{Frequency Range 2}
\newacronym{fs}{FS}{Fast Switching}
\newacronym{fscc}{FSCC}{Flow Sharing Congestion Control}
\newacronym{ftp}{FTP}{File Transfer Protocol}
\newacronym{fw}{FW}{Flow Window}
\newacronym{ge}{GE}{Gaussian Elimination}
\newacronym{gnb}{gNB}{Next Generation Node Base}
\newacronym{gop}{GOP}{Group of Pictures}
\newacronym{gpr}{GPR}{Gaussian Process Regressor}
\newacronym{gpu}{GPU}{Graphics Processing Unit}
\newacronym{gtp}{GTP}{GPRS Tunneling Protocol}
\newacronym{gtpc}{GTP-C}{GPRS Tunnelling Protocol Control Plane}
\newacronym{gtpu}{GTP-U}{GPRS Tunnelling Protocol User Plane}
\newacronym{gtpv2c}{GTPv2-C}{\gls{gtp} v2 - Control}
\newacronym{gw}{GW}{Gateway}
\newacronym{harq}{HARQ}{Hybrid Automatic Repeat reQuest}
\newacronym{hetnet}{HetNet}{Heterogeneous Network}
\newacronym{hh}{HH}{Hard Handover}
\newacronym{hol}{HOL}{Head-of-Line}
\newacronym{hqf}{HQF}{Highest-quality-first}
\newacronym{hss}{HSS}{Home Subscription Server}
\newacronym{http}{HTTP}{HyperText Transfer Protocol}
\newacronym{ia}{IA}{Initial Access}
\newacronym{iab}{IAB}{Integrated Access and Backhaul}
\newacronym{ic}{IC}{Incident Command}
\newacronym{ietf}{IETF}{Internet Engineering Task Force}
\newacronym{imsi}{IMSI}{International Mobile Subscriber Identity}
\newacronym{imt}{IMT}{International Mobile Telecommunication}
\newacronym{iot}{IoT}{Internet of Things}
\newacronym{ip}{IP}{Internet Protocol}
\newacronym{itu}{ITU}{International Telecommunication Union}
\newacronym{kpi}{KPI}{Key Performance Indicator}
\newacronym{kpm}{KPM}{Key Performance Measurement}
\newacronym{kvm}{KVM}{Kernel-based Virtual Machine}
\newacronym{los}{LOS}{Line-of-Sight}
\newacronym{lsm}{LSM}{Link-to-System Mapping}
\newacronym{lstm}{LSTM}{Long Short Term Memory}
\newacronym{lte}{LTE}{Long Term Evolution}
\newacronym{lxc}{LXC}{Linux Container}
\newacronym{m2m}{M2M}{Machine to Machine}
\newacronym{mac}{MAC}{Medium Access Control}
\newacronym{manet}{MANET}{Mobile Ad Hoc Network}
\newacronym{mano}{MANO}{Management and Orchestration}
\newacronym{mc}{MC}{Multi-Connectivity}
\newacronym{mcc}{MCC}{Mobile Cloud Computing}
\newacronym{mchem}{MCHEM}{Massive Channel Emulator}
\newacronym{mcs}{MCS}{Modulation and Coding Scheme}
\newacronym{mec}{MEC}{Multi-access Edge Computing}
\newacronym{mec2}{MEC}{Mobile Edge Cloud}
\newacronym{mfc}{MFC}{Mobile Fog Computing}
\newacronym{mgen}{MGEN}{Multi-Generator}
\newacronym{mi}{MI}{Mutual Information}
\newacronym{mib}{MIB}{Master Information Block}
\newacronym{miesm}{MIESM}{Mutual Information Based Effective SINR}
\newacronym{mimo}{MIMO}{Multiple Input, Multiple Output}
\newacronym{ml}{ML}{Machine Learning}
\newacronym{mlr}{MLR}{Maximum-local-rate}
\newacronym[plural=\gls{mme}s,firstplural=Mobility Management Entities (MMEs)]{mme}{MME}{Mobility Management Entity}
\newacronym{mmtc}{mMTC}{Massive Machine-Type Communications}
\newacronym{mmwave}{mmWave}{millimeter wave}
\newacronym{mpdccp}{MP-DCCP}{Multipath Datagram Congestion Control Protocol}
\newacronym{mptcp}{MPTCP}{Multipath TCP}
\newacronym{mr}{MR}{Maximum Rate}
\newacronym{mrdc}{MR-DC}{Multi \gls{rat} \gls{dc}}
\newacronym{mse}{MSE}{Mean Square Error}
\newacronym{mss}{MSS}{Maximum Segment Size}
\newacronym{mt}{MT}{Mobile Termination}
\newacronym{mtd}{MTD}{Machine-Type Device}
\newacronym{mtu}{MTU}{Maximum Transmission Unit}
\newacronym{mumimo}{MU-MIMO}{Multi-user \gls{mimo}}
\newacronym{mvno}{MVNO}{Mobile Virtual Network Operator}
\newacronym{nalu}{NALU}{Network Abstraction Layer Unit}
\newacronym{nas}{NAS}{Network Attached Storage}
\newacronym{nbiot}{NB-IoT}{Narrow Band IoT}
\newacronym{nfv}{NFV}{Network Function Virtualization}
\newacronym{nfvi}{NFVI}{Network Function Virtualization Infrastructure}
\newacronym{ni}{NI}{Network Interfaces}
\newacronym{nic}{NIC}{Network Interface Card}
\newacronym{nlos}{NLOS}{Non-Line-of-Sight}
\newacronym{now}{NOW}{Non Overlapping Window}
\newacronym{nsm}{NSM}{Network Service Mesh}
\newacronym[type=hidden]{nr}{NR}{New Radio}
\newacronym{nrf}{NRF}{Network Repository Function}
\newacronym{nsa}{NSA}{Non Stand Alone}
\newacronym{nse}{NSE}{Network Slicing Engine}
\newacronym{nssf}{NSSF}{Network Slice Selection Function}
\newacronym{o2i}{O2I}{Outdoor to Indoor}
\newacronym{oai}{OAI}{OpenAirInterface}
\newacronym{oaicn}{OAI-CN}{\gls{oai} \acrlong{cn}}
\newacronym{oairan}{OAI-RAN}{\acrlong{oai} \acrlong{ran}}
\newacronym{oam}{OAM}{Operations, Administration and Maintenance}
\newacronym{ofdm}{OFDM}{Orthogonal Frequency Division Multiplexing}
\newacronym{olia}{OLIA}{Opportunistic Linked Increase Algorithm}
\newacronym{omec}{OMEC}{Open Mobile Evolved Core}
\newacronym{onap}{ONAP}{Open Network Automation Platform}
\newacronym{onf}{ONF}{Open Networking Foundation}
\newacronym{onos}{ONOS}{Open Networking Operating System}
\newacronym{oom}{OOM}{\gls{onap} Operations Manager}
\newacronym{opnfv}{OPNFV}{Open Platform for \gls{nfv}}
\newacronym[type=hidden]{oran}{O-RAN}{Open \gls{ran}}
\newacronym{orbit}{ORBIT}{Open-Access Research Testbed for Next-Generation Wireless Networks}
\newacronym{os}{OS}{Operating System}
\newacronym{oss}{OSS}{Operations Support System}
\newacronym{pa}{PA}{Position-aware}
\newacronym{pase}{PASE}{Prioritization, Arbitration, and Self-adjusting Endpoints}
\newacronym{pawr}{PAWR}{Platforms for Advanced Wireless Research}
\newacronym{pbch}{PBCH}{Physical Broadcast Channel}
\newacronym{pcef}{PCEF}{Policy and Charging Enforcement Function}
\newacronym{pcfich}{PCFICH}{Physical Control Format Indicator Channel}
\newacronym{pcrf}{PCRF}{Policy and Charging Rules Function}
\newacronym{pdcch}{PDCCH}{Physical Downlink Control Channel}
\newacronym{pdcp}{PDCP}{Packet Data Convergence Protocol}
\newacronym{pdsch}{PDSCH}{Physical Downlink Shared Channel}
\newacronym{pdu}{PDU}{Packet Data Unit}
\newacronym{pf}{PF}{Proportional Fair}
\newacronym{pgw}{PGW}{Packet Gateway}
\newacronym{phich}{PHICH}{Physical Hybrid ARQ Indicator Channel}
\newacronym{phy}{PHY}{Physical}
\newacronym{pmch}{PMCH}{Physical Multicast Channel}
\newacronym{pmi}{PMI}{Precoding Matrix Indicators}
\newacronym{powder}{POWDER}{Platform for Open Wireless Data-driven Experimental Research}
\newacronym{ppo}{PPO}{Proximal Policy Optimization}
\newacronym{ppp}{PPP}{Poisson Point Process}
\newacronym{prach}{PRACH}{Physical Random Access Channel}
\newacronym{prb}{PRB}{Physical Resource Block}
\newacronym{psnr}{PSNR}{Peak Signal to Noise Ratio}
\newacronym{pss}{PSS}{Primary Synchronization Signal}
\newacronym{pucch}{PUCCH}{Physical Uplink Control Channel}
\newacronym{pusch}{PUSCH}{Physical Uplink Shared Channel}
\newacronym{qam}{QAM}{Quadrature Amplitude Modulation}
\newacronym{qci}{QCI}{\gls{qos} Class Identifier}
\newacronym{qoe}{QoE}{Quality of Experience}
\newacronym{qos}{QoS}{Quality of Service}
\newacronym{quic}{QUIC}{Quick UDP Internet Connections}
\newacronym{rach}{RACH}{Random Access Channel}
\newacronym{ran}{RAN}{Radio Access Network}
\newacronym[firstplural=Radio Access Technologies (RATs)]{rat}{RAT}{Radio Access Technology}
\newacronym{rbg}{RBG}{Resource Block Group}
\newacronym{rcn}{RCN}{Research Coordination Network}
\newacronym{rc}{RC}{RAN Control}
\newacronym{rec}{REC}{Radio Edge Cloud}
\newacronym{red}{RED}{Random Early Detection}
\newacronym{renew}{RENEW}{Reconfigurable Eco-system for Next-generation End-to-end Wireless}
\newacronym{rf}{RF}{Radio Frequency}
\newacronym{rfc}{RFC}{Request for Comments}
\newacronym{rfr}{RFR}{Random Forest Regressor}
\newacronym{ric}{RIC}{\gls{ran} Intelligent Controller}
\newacronym{rlc}{RLC}{Radio Link Control}
\newacronym{rlf}{RLF}{Radio Link Failure}
\newacronym{rlnc}{RLNC}{Random Linear Network Coding}
\newacronym{rmr}{RMR}{RIC Message Router}
\newacronym{rmse}{RMSE}{Root Mean Squared Error}
\newacronym{rnis}{RNIS}{Radio Network Information Service}
\newacronym{rr}{RR}{Round Robin}
\newacronym{rrc}{RRC}{Radio Resource Control}
\newacronym{rrm}{RRM}{Radio Resource Management}
\newacronym{rru}{RRU}{Remote Radio Unit}
\newacronym{rs}{RS}{Remote Server}
\newacronym{rsrp}{RSRP}{Reference Signal Received Power}
\newacronym{rsrq}{RSRQ}{Reference Signal Received Quality}
\newacronym{rss}{RSS}{Received Signal Strength}
\newacronym{rssi}{RSSI}{Received Signal Strength Indicator}
\newacronym{rtt}{RTT}{Round Trip Time}
\newacronym{ru}{RU}{Radio Unit}
\newacronym{rw}{RW}{Receive Window}
\newacronym{rx}{RX}{Receiver}
\newacronym{s1ap}{S1AP}{S1 Application Protocol}
\newacronym{sa}{SA}{standalone}
\newacronym{sack}{SACK}{Selective Acknowledgment}
\newacronym{sap}{SAP}{Service Access Point}
\newacronym{sc2}{SC2}{Spectrum Collaboration Challenge}
\newacronym{scef}{SCEF}{Service Capability Exposure Function}
\newacronym{sch}{SCH}{Secondary Cell Handover}
\newacronym{scoot}{SCOOT}{Split Cycle Offset Optimization Technique}
\newacronym{sctp}{SCTP}{Stream Control Transmission Protocol}
\newacronym{sdap}{SDAP}{Service Data Adaptation Protocol}
\newacronym{sdk}{SDK}{Software Development Kit}
\newacronym{sdm}{SDM}{Space Division Multiplexing}
\newacronym{sdma}{SDMA}{Spatial Division Multiple Access}
\newacronym{sdn}{SDN}{Software-defined Networking}
\newacronym{sdr}{SDR}{Software-defined Radio}
\newacronym{seba}{SEBA}{SDN-Enabled Broadband Access}
\newacronym{sgsn}{SGSN}{Serving GPRS Support Node}
\newacronym{sgw}{SGW}{Service Gateway}
\newacronym{si}{SI}{Study Item}
\newacronym{sib}{SIB}{Secondary Information Block}
\newacronym{sinr}{SINR}{Signal to Interference plus Noise Ratio}
\newacronym{sip}{SIP}{Session Initiation Protocol}
\newacronym{siso}{SISO}{Single Input, Single Output}
\newacronym{sla}{SLA}{Service Level Agreement}
\newacronym{sm}{SM}{Service Model}
\newacronym{smf}{SMF}{Session Management Function}
\newacronym{smo}{SMO}{Service Management and Orchestration}
\newacronym{sms}{SMS}{Short Message Service}
\newacronym{smsgmsc}{SMS-GMSC}{\gls{sms}-Gateway}
\newacronym{snr}{SNR}{Signal-to-Noise-Ratio}
\newacronym{son}{SON}{Self-Organizing Network}
\newacronym{sptcp}{SPTCP}{Single Path TCP}
\newacronym{srb}{SRB}{Service Radio Bearer}
\newacronym{srn}{SRN}{Standard Radio Node}
\newacronym{srs}{SRS}{Sounding Reference Signal}
\newacronym{ss}{SS}{Synchronization Signal}
\newacronym{sss}{SSS}{Secondary Synchronization Signal}
\newacronym{st}{ST}{Spanning Tree}
\newacronym{svc}{SVC}{Scalable Video Coding}
\newacronym{tb}{TB}{Transport Block}
\newacronym{tcp}{TCP}{Transmission Control Protocol}
\newacronym{tdd}{TDD}{Time Division Duplexing}
\newacronym{tdm}{TDM}{Time Division Multiplexing}
\newacronym{tdma}{TDMA}{Time Division Multiple Access}
\newacronym{tfl}{TfL}{Transport for London}
\newacronym{tfrc}{TFRC}{TCP-Friendly Rate Control}
\newacronym{tft}{TFT}{Traffic Flow Template}
\newacronym{tgen}{TGEN}{Traffic Generator}
\newacronym{tip}{TIP}{Telecom Infra Project}
\newacronym{tm}{TM}{Transparent Mode}
\newacronym{to}{TO}{Telco Operator}
\newacronym{tr}{TR}{Technical Report}
\newacronym{trp}{TRP}{Transmitter Receiver Pair}
\newacronym{ts}{TS}{Technical Specification}
\newacronym{tti}{TTI}{Transmission Time Interval}
\newacronym{ttt}{TTT}{Time-to-Trigger}
\newacronym{tx}{TX}{Transmitter}
\newacronym{uas}{UAS}{Unmanned Aerial System}
\newacronym{uav}{UAV}{Unmanned Aerial Vehicle}
\newacronym{udm}{UDM}{Unified Data Management}
\newacronym{udp}{UDP}{User Datagram Protocol}
\newacronym{udr}{UDR}{Unified Data Repository}
\newacronym{ue}{UE}{User Equipment}
\newacronym{uhd}{UHD}{\gls{usrp} Hardware Driver}
\newacronym{ul}{UL}{Uplink}
\newacronym{um}{UM}{Unacknowledged Mode}
\newacronym{uml}{UML}{Unified Modeling Language}
\newacronym{upa}{UPA}{Uniform Planar Array}
\newacronym{upf}{UPF}{User Plane Function}
\newacronym{urllc}{URLLC}{Ultra Reliable and Low Latency Communications}
\newacronym{usa}{U.S.}{United States}
\newacronym{usim}{USIM}{Universal Subscriber Identity Module}
\newacronym{usrp}{USRP}{Universal Software Radio Peripheral}
\newacronym{utc}{UTC}{Urban Traffic Control}
\newacronym{vim}{VIM}{Virtualization Infrastructure Manager}
\newacronym{vm}{VM}{Virtual Machine}
\newacronym{vnf}{VNF}{Virtual Network Function}
\newacronym{volte}{VoLTE}{Voice over \gls{lte}}
\newacronym{voltha}{VOLTHA}{Virtual OLT HArdware Abstraction}
\newacronym{vr}{VR}{Virtual Reality}
\newacronym{vran}{vRAN}{Virtualized \gls{ran}}
\newacronym{vss}{VSS}{Video Streaming Server}
\newacronym{wbf}{WBF}{Wired Bias Function}
\newacronym{wf}{WF}{Waterfilling}
\newacronym{wg}{WG}{Working Group}
\newacronym{wlan}{WLAN}{Wireless Local Area Network}
\newacronym{osm}{OSM}{Open Source \gls{nfv} Management and Orchestration}
\newacronym{pnf}{PNF}{Physical Network Function}
\newacronym{drl}{DRL}{Deep Reinforcement Learning}
\newacronym{mtc}{MTC}{Machine-type Communications}
\newacronym{osc}{OSC}{O-RAN Software Community}
\newacronym{mns}{MnS}{Management Services}
\newacronym{ves}{VES}{\gls{vnf} Event Stream}
\newacronym{ei}{EI}{Enrichment Information}
\newacronym{fh}{FH}{Fronthaul}
\newacronym{fft}{FFT}{Fast Fourier Transform}
\newacronym{laa}{LAA}{Licensed-Assisted Access}
\newacronym{plfs}{PLFS}{Physical Layer Frequency Signals}
\newacronym{ptp}{PTP}{Precision Time Protocol}
\tikzstyle{startstop} = [rectangle, rounded corners, minimum width=2cm, minimum height=0.5cm,text centered, draw=black]
\tikzstyle{io} = [trapezium, trapezium left angle=70, trapezium right angle=110, minimum width=3cm, minimum height=1cm, text centered, draw=black]
\tikzstyle{process} = [rectangle, minimum width=2cm, minimum height=0.5cm, text centered, draw=black, alignb=center]
\tikzstyle{decision} = [ellipse, minimum width=2cm, minimum height=1cm, text centered, draw=black]
\tikzstyle{arrow} = [thick,<->,>=stealth]
\tikzstyle{line} = [thick,>=stealth]
\tikzstyle{darrow} = [thick,<->,>=stealth,dashed]
\tikzstyle{sarrow} = [thick,->,>=stealth]
\tikzstyle{larrow} = [line width=0.1mm,dashdotted,->,>=stealth]
\tikzstyle{llarrow} = [line width=0.1mm,->,>=stealth]

\makeatletter
\def\grd@save@target#1{%
  \def\grd@target{#1}}
\def\grd@save@start#1{%
  \def\grd@start{#1}}
\tikzset{
  grid with coordinates/.style={
    to path={%
      \pgfextra{%
        \edef\grd@@target{(\tikztotarget)}%
        \tikz@scan@one@point\grd@save@target\grd@@target\relax
        \edef\grd@@start{(\tikztostart)}%
        \tikz@scan@one@point\grd@save@start\grd@@start\relax
        \draw[minor help lines] (\tikztostart) grid (\tikztotarget);
        \draw[major help lines] (\tikztostart) grid (\tikztotarget);
        \grd@start
        \pgfmathsetmacro{\grd@xa}{\the\pgf@x/1cm}
        \pgfmathsetmacro{\grd@ya}{\the\pgf@y/1cm}
        \grd@target
        \pgfmathsetmacro{\grd@xb}{\the\pgf@x/1cm}
        \pgfmathsetmacro{\grd@yb}{\the\pgf@y/1cm}
        \pgfmathsetmacro{\grd@xc}{\grd@xa + \pgfkeysvalueof{/tikz/grid with coordinates/major step x}}
        \pgfmathsetmacro{\grd@yc}{\grd@ya + \pgfkeysvalueof{/tikz/grid with coordinates/major step y}}
        \foreach \x in {\grd@xa,\grd@xc,...,\grd@xb}
        \node[anchor=north] at (\x,\grd@ya) {\pgfmathprintnumber{\x}};
        \foreach \y in {\grd@ya,\grd@yc,...,\grd@yb}
        \node[anchor=east] at (\grd@xa,\y) {\pgfmathprintnumber{\y}};
      }
    }
  },
  minor help lines/.style={
    help lines,
    gray,
    line cap =round,
    xstep=\pgfkeysvalueof{/tikz/grid with coordinates/minor step x},
    ystep=\pgfkeysvalueof{/tikz/grid with coordinates/minor step y}
  },
  major help lines/.style={
    help lines,
    line cap =round,
    line width=\pgfkeysvalueof{/tikz/grid with coordinates/major line width},
    xstep=\pgfkeysvalueof{/tikz/grid with coordinates/major step x},
    ystep=\pgfkeysvalueof{/tikz/grid with coordinates/major step y}
  },
  grid with coordinates/.cd,
  minor step x/.initial=.5,
  minor step y/.initial=.2,
  major step x/.initial=1,
  major step y/.initial=1,
  major line width/.initial=1pt,
}
\makeatother

\definecolor{desireRed}{RGB}{230,57,60}%
\definecolor{darkPurple}{RGB}{59,31,43}%
\definecolor{springGreen}{RGB}{37,223,145}%
\definecolor{queenBlue}{RGB}{69,123,157}%
\definecolor{spaceCadet}{RGB}{29,53,87}%

\usepackage{dblfloatfix}

\newcommand{\coloran}{ColO-RAN\xspace}
\newcommand{\openrangym}{OpenRAN Gym\xspace}
\newcommand{\scope}{SCOPE\xspace}

\usepackage{enumitem}
\setlist[itemize]{align=parleft,left=0pt..1em,wide=0pt,topsep=0pt,partopsep=0pt}

\usepackage[var0]{inconsolata}

\IEEEoverridecommandlockouts

\begin{document}

\title{OpenRAN Gym: An Open Toolbox for Data Collection and Experimentation with AI in O-RAN}

\author{\IEEEauthorblockN{Leonardo Bonati, Michele Polese, Salvatore D'Oro, Stefano Basagni, Tommaso Melodia}
\IEEEauthorblockA{Institute for the Wireless Internet of Things, Northeastern University, Boston, MA, U.S.A.\\
E-mail: \{l.bonati, m.polese, s.doro, s.basagni, melodia\}@northeastern.edu}
\thanks{This work was partially supported by the U.S.\ National Science Foundation under Grants CNS-1925601, CNS-2120447, and CNS-2112471.}
}

\makeatletter
\patchcmd{\@maketitle}
  {\addvspace{0.5\baselineskip}\egroup}
  {\addvspace{-1.5\baselineskip}\egroup}
  {}
  {}
\makeatother

\flushbottom
\setlength{\parskip}{0ex plus0.1ex}

\maketitle
\glsunset{nr}
\glsunset{lte}
\glsunset{3gpp}
\glsunset{usrp}
\glsunset{fpga}
\glsunset{gpu}

\begin{abstract}
Open \gls{ran} architectures will enable interoperability, openness, and programmatic data-driven control in next generation cellular networks. 
However, developing scalable and efficient data-driven algorithms that can generalize across diverse deployments and optimize \gls{ran} performance is a complex feat, largely unaddressed as of today. 
Specifically, the ability to design efficient data-driven algorithms for network control and inference requires at a minimum (i) access to large, rich, and heterogeneous datasets; (ii) testing at scale in controlled but realistic environments, and (iii) software pipelines to automate data collection and experimentation.
To facilitate these tasks, in this paper we propose \openrangym, a practical, open,  experimental toolbox that provides end-to-end design, data collection, and testing workflows for intelligent control in next generation Open \gls{ran} systems. 
\openrangym builds on software frameworks for the collection of large datasets and \gls{ran} control, and on a lightweight O-RAN environment for experimental wireless platforms.
We first provide an overview of \openrangym and  then describe how it can be used to collect data, to design and train artificial intelligence and machine learning-based O-RAN applications (xApps), and to test xApps on a softwarized \gls{ran}.
Then, we provide an example of two xApps designed with \openrangym and used to control a large-scale network with~7 base stations and~42 users deployed on the Colosseum testbed.
\openrangym and its software components are open source and
publicly-available to the research community.
\end{abstract}

\begin{picture}(0,0)(10,-410)
\put(0,0){
\put(0,10){\footnotesize This paper has been accepted for publication on IEEE WCNC 2022 Workshop on Open RAN Architecture for 5G Evolution and 6G.}
\put(0,0){\tiny \copyright 2022 IEEE. Personal use of this material is permitted. Permission from IEEE must be obtained for all other uses, in any current or future media including reprinting/republishing}
\put(0,-5){\tiny this material for advertising or promotional purposes, creating new collective works, for resale or redistribution to servers or lists, or reuse of any copyrighted component of this work in other works.}
\put(0,-20){\scriptsize }}
\end{picture}

\glsresetall
\glsunset{nr}
\glsunset{lte}
\glsunset{3gpp}
\glsunset{usrp}
\glsunset{fpga}
\glsunset{gpu}

\vspace{-10pt}
\section{Introduction}

The next generations of cellular networks will follow the Open \gls{ran} paradigm, which promotes openness, virtualization, programmability, and data-driven control loops in the mobile environment~\cite{bonati2020open}. This will help network operators support new bespoke services on the same physical infrastructure, thanks to the flexibility and reconfigurability
of software-based deployments and algorithmic control. Open \gls{ran} will also decrease operational costs because of the increased efficiency enabled by virtualized and open ecosystems.

In this context, the O-RAN Alliance has developed specifications to apply the Open RAN paradigm to prevailing radio access technologies including \gls{3gpp} LTE and NR networks~\cite{oran-wg1-arch-spec}. 
O-RAN specifications introduce standardized interfaces that connect new, O-RAN-specific network nodes to key \gls{ran} elements, such as the NR \glspl{cu}, \glspl{du}, \glspl{ru}, and the LTE O-RAN-compliant \glspl{enb}~\cite{polese2022understanding}.
To enable programmatic closed-loop control of the \gls{ran}, O-RAN also introduces two so-called \glspl{ric}.
The near-real-time (or near-RT) \gls{ric} is connected through the E2 interface to the \gls{ran} network functions (i.e. the \gls{cu} and the \gls{du}) to enable control loops that operate at timescales between~$10$\:ms and~$1$\:s~\cite{oran-wg3-ricarch}.
%
The non-real-time (non-RT) \gls{ric}, instead, is part of the service management and orchestration framework and operates at timescales larger than $1$\:s~\cite{oran-wg2-non-rt-ric-architecture}. 
This component connects to one or multiple near-RT \glspl{ric} through the A1 interface, used to distribute policies, external information, and to manage \gls{ai} and \gls{ml} models.
These models define intelligent network control strategies that are then run on the \glspl{ric} in the form of applications---namely, xApps and rApps---that can be provided by \gls{ran} vendors, operators or third-party entities.
xApps run in the near-RT \gls{ric}, while rApps run in the non-RT \gls{ric}.
%

The open and disaggregated O-RAN architecture enables the practical deployment of 
\gls{ai}/\gls{ml} solutions at scale.
For instance, \gls{ai}/\gls{ml} algorithms can perform inference and traffic forecasting or configure \gls{ran} nodes based on run-time conditions and traffic requirements.
The O-RAN specifications~\cite{oran-wg2-ml} discuss the typical  workflow for the development and testing of \gls{ai}/\gls{ml} in the \gls{ran}~\cite{polese2021coloran}.
This involves multiple steps, including:
(i)~\emph{data collection}, to create rich datasets to capture the characteristics of the environment where the \gls{ai}/\gls{ml} solution will be deployed (e.g., wireless channel,
user distributions and requirements) as well as various indicators of network performance under different configurations;
(ii)~\emph{AI/ML model design}, with a selection of the model inputs and outputs, and \emph{training and testing}, to understand its limits and effectiveness;
(iii)~\emph{model deployment} as xApps or rApps;
(iv)~\emph{model fine-tuning} with live data from the \gls{ran}, to adapt it to the production environment,
and, finally, (v)~the actual \emph{inference, forecasting and/or control} of the \gls{ran}. 

\begin{figure*}
\setlength\belowcaptionskip{-10pt}
    \centering
    \includegraphics[width=.9\textwidth]{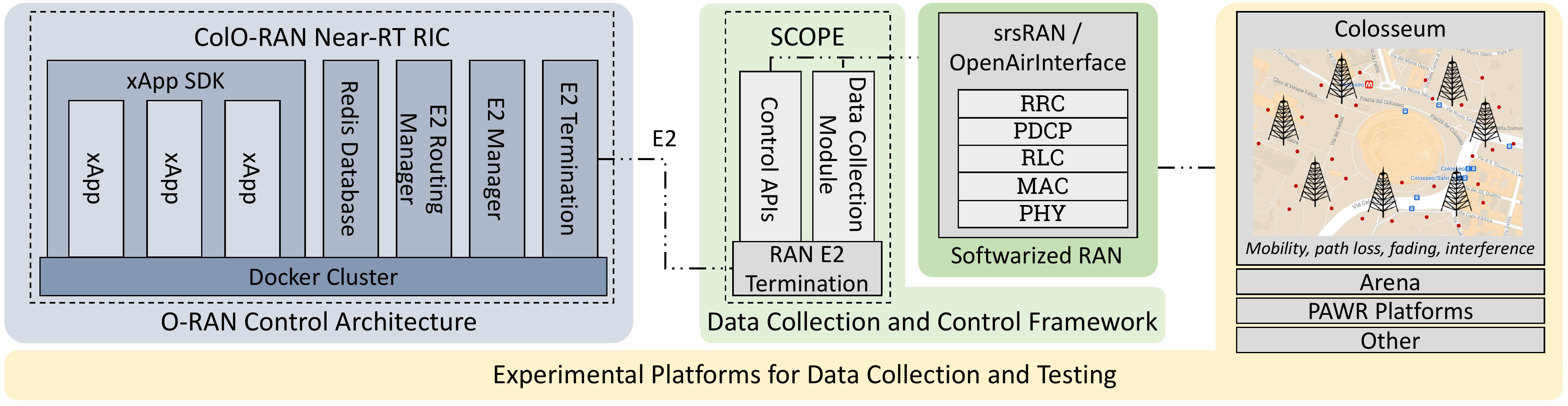}
    \caption{\openrangym architecture.}
    \label{fig:openran-architecture}
\end{figure*}

In this paper, we introduce \openrangym, an open toolbox to develop O-RAN-compliant \gls{ai}/\gls{ml} solutions, deploy them as xApps on the near-RT \gls{ric}, and test them on large-scale softwarized \glspl{ran} controlled by the \gls{ric}.
We first give an overview of \openrangym and its core components, discussing how they enable design and testing workflows of \gls{ml}-based xApps.
Then, we demonstrate how two xApps designed with \openrangym can be used to control a large-scale \gls{ran} instantiated on the Colosseum wireless network emulator through the \scope framework~\cite{bonati2021scope}, and managed by an O-RAN near-RT \gls{ric} provided by the \coloran framework~\cite{polese2021coloran}.
To the best of our knowledge, this is the first open toolset for the end-to-end design and experimentation of data-driven O-RAN xApps on large-scale experimental platforms.
%
%

Previous experimental work has focused on the development of data-driven solutions and xApps for specific use cases~\cite{s21248173,johnson2021open}, on the description of the \gls{ai}/\gls{ml} capabilities of O-RAN~\cite{lee2020hosting,abdalla2021generation}, on interoperability testing~\cite{oran2019plugfest}, and on orchestration~\cite{doro2022orchestran}. Compared to the state of the art, \openrangym enables an end-to-end workflow for the design and testing of \gls{ai}/\gls{ml} solutions as xApps in the O-RAN ecosystem.
By doing so, it empowers users with a first-of-its-kind open and publicly-available O-RAN-compliant toolbox that will unleash the potential of data-driven applications for next generation cellular networks.
\openrangym aims at creating a thriving community of researchers and developers contributing to it with open source software components for experimental O-RAN-driven \gls{ai}/\gls{ml} research.\footnote{The software components of \openrangym are publicly-available and accessible at \url{https://openrangym.com}.}
%

The remainder of this paper is organized as follows. 
Section~\ref{sec:openrangym} provides an overview of \openrangym.
A practical description of the \scope and \coloran frameworks is given in Sections~\ref{sec:scope} and~\ref{sec:coloran}. 
Section~\ref{sec:xapp-design-testing-workflow} presents the xApp design and testing workflow, and provides an example of large-scale \gls{ran} control using xApps developed with \openrangym on Colosseum. 
Finally, Section~\ref{sec:conclusions} concludes the paper.

\section{OpenRAN Gym}
\label{sec:openrangym}

The architecture of \openrangym is shown in Fig.~\ref{fig:openran-architecture}.
It includes: (i)~one or multiple publicly-accessible \textit{experimental platforms} for data collection and testing (e.g., Colosseum~\cite{bonati2021colosseum}, Arena~\cite{bertizzolo2020arena}, the platforms of the PAWR program~\cite{pawr}); (ii)~a \textit{softwarized \gls{ran}} (e.g., implemented through srsRAN~\cite{gomez2016srslte} or OpenAirInterface~\cite{kaltenberger2020openairinterface}); (iii)~a \textit{data collection and control framework} with \glspl{api} to control the cellular stacks and extract statistics from them (e.g., \scope~\cite{bonati2021scope}), and (iv)~an \textit{O-RAN control architecture} with interfaces to connect to the \gls{ran} and control it through \gls{ai}/\gls{ml} solutions (e.g., \coloran~\cite{polese2021coloran}).
The platform-independence of \openrangym allows users to collect data, design and train solutions in heterogeneous environments before deploying them on production networks.
In this way, several evaluation and fine-tuning iterations can be performed in controlled setups to guarantee that the final \gls{ai}/\gls{ml} model behaves as expected.

At the time of this writing, \openrangym supports Arena~\cite{bertizzolo2020arena} and Colosseum~\cite{bonati2021colosseum} as experimental platforms.
Arena is an indoor testbed that includes a grid of 64~indoor antennas and 24~\glspl{sdr} controlled by dedicated compute servers.
Colosseum is a large-scale wireless network emulator that allows users to test solutions at scale through 128~\gls{usrp} X310 \glspl{sdr} and compute servers (called \glspl{srn}), and to emulate the conditions of wireless environments (e.g., path loss, fading, user mobility and signal interference) through a \gls{mchem} that controls 128~additional \gls{usrp} X310.
\gls{mchem} is capable of capturing the conditions of the wireless channel with high fidelity. Wireless channels can be modeled through ray-tracing software~\cite{tehranimoayyed2021creating}, reproduced in \gls{mchem} through \gls{fpga}-based finite impulse response filters.
In this way, Colosseum allows users to perform experiments in a multitude of different emulated terrains, as if the \glspl{sdr} were operating in the real-world.
Similarly, Colosseum can also emulate traffic with different characteristics and distributions through the \gls{tgen} system,
which generates and manages traffic flows between the \glspl{srn}.
%

In the current version of \openrangym, the softwarized \gls{ran} is based on srsRAN~\cite{gomez2016srslte}, which implements the full stack of a \gls{3gpp} \gls{enb} and controls the \glspl{usrp} X310 of the \glspl{srn} that act as radio front-ends.
The srsRAN protocol stack is extended by the \scope framework, which enhances it with additional networking and control functionalities.
%
These include \gls{ran} slicing capabilities, additional scheduling policies, and data collection functionalities, as well as novel \glspl{api} to control such capabilities at run time.
%
%
%
As we will discuss in Section~\ref{sec:scope}, this component leverages the emulation capabilities of Colosseum---which acts as a \textit{wireless data factory}---to enable automated, large-scale data collection campaigns to create datasets with tens of hours of \gls{ran} experiments and in different wireless and traffic conditions~\cite{bonati2021intelligence,polese2021coloran,casasole2021qcell}.

Finally, the O-RAN control architecture is implemented through \coloran, which extends and adapts the \gls{osc} near-RT \gls{ric} to the Colosseum environment, and connects \scope-enabled base stations to the near-RT \gls{ric} through the O-RAN E2 interface.
%
As we will discuss in Section~\ref{sec:coloran}, \coloran provides an instance of an O-RAN-compliant near-RT \gls{ric}, together with an \textit{xApp \gls{sdk}} that allows users to design and test \gls{ai}/\gls{ml}-based xApps (Fig.~\ref{fig:openran-architecture}).
%



\section{Data Collection and Control Framework}
\label{sec:scope}

\openrangym leverages \scope~\cite{bonati2021scope} as the data collection and control framework.
\scope provides a development environment to design, prototype and test adaptive solutions for cellular networking, and for  large-scale data collection of \gls{ran} \glspl{kpm}.
It builds on top of srsRAN~\cite{gomez2016srslte}, and extends it with novel functionalities (e.g., network slicing and additional scheduling policies), open \glspl{api} to control the \gls{ran} configuration at run time (e.g., the resources allocated to each slice), and data collection capabilities.
 \scope has powerful data collection capabilities when used in combination with Colosseum, which makes it possible to automatically perform large-scale data collection campaigns in realistic wireless environments~\cite{bonati2021scope}.
%
Examples include data collection campaigns with tens of hours of experiments, and in setups with up to~49 nodes (7~base stations and 42~users) in sliced \glspl{ran} with different \gls{qos} requirements, scheduling policies, slicing resources, and in different emulation scenarios~\cite{bonati2021intelligence,polese2021coloran,casasole2021qcell}.
%
%
\scope has been extended to incorporate a \gls{ran}-side E2 termination (based on the \gls{osc} \gls{du}~\cite{osc-du-l2}) to connect to the O-RAN near-RT \gls{ric}.
This allows xApps running on the \gls{ric} to interface with the \scope \glspl{api} and control the functionalities of the base stations at run time.

In the remaining of this section, we will show how to configure the main parameters of the \scope base stations, and how to start \scope on Colosseum. 
It is worth mentioning that even if we specifically focus on the implementation for the Colosseum network emulator (provided as a ready-to-use container image, namely \texttt{scope-e2}), \scope can be ported to different platforms with minor adaptations (see~\cite[Section~5.3]{bonati2021scope}).

\subsection{Starting  \scope}
\label{sec:scope-setup}

\begin{table}[t]
\setlength\belowcaptionskip{5pt}
    \centering
    \footnotesize
    \setlength{\tabcolsep}{2pt}
    \caption{Main \scope slicing and scheduling configuration parameters.}
    \label{tab:scope-parameters}
    \begin{tabularx}{\columnwidth}{
        >{\raggedright\arraybackslash\hsize=0.95\hsize}X 
        >{\raggedright\arraybackslash\hsize=1.05\hsize}X }
        \toprule
        Name & Description \\
        \midrule
        \texttt{network-slicing} & Enable/disable network slicing \\
        \texttt{slice-allocation} & Define the base station slice allocation \\
        \texttt{slice-scheduling-policy} & Set the scheduling policy for each slice \\
        \texttt{slice-users} & Assign UEs to slices \\
        \bottomrule
    \end{tabularx}
\vspace{-12pt}
\end{table}

\scope allows users to configure the base stations through JSON files.
The main parameters to set up the network slicing and scheduling functionalities (see Table~\ref{tab:scope-parameters}) are as follows.\footnote{A comprehensive description of all parameters of the \scope APIs configuration files can be found at \url{https://github.com/wineslab/colosseum-scope}.}

\begin{itemize}
\item \texttt{network-slicing}: this parameter enables/disables the network slicing capabilities of the \scope base station.

\item \texttt{slice-allocation}: sets the \glspl{rbg} of each slice.
It takes as input \texttt{\{slice:[first\_rbg, last\_rbg],...\}}, e.g., \texttt{\{0:[0,3],1:[5,7]\}} assigns \glspl{rbg}~0-3 to slice~0 and \glspl{rbg}~5-7 to slice~1.

\item \texttt{slice-scheduling-policy}: sets the scheduling policy the base station uses for each slice, e.g., \texttt{[2,0]} assigns policy~2 to slice~0 and policy~0 to slice~1.
The possible values correspond to the scheduling policies supported by \scope (0~for round-robin, 1~for waterfilling, 2~for proportionally fair).
%

\item \texttt{slice-users}: assigns \glspl{ue} to the slices.
It takes as input \texttt{\{slice:[ue1,ue2],...\}}, e.g., \texttt{\{0:[2,5],1:[3,4]\}} assigns \glspl{ue}~2, 5 to slice~0, and \glspl{ue}~3, 4 to slice~1.
The \gls{ue} ID corresponds to the $i$-th \gls{srn} allocated to the experimenter (the base station is assumed as the first \gls{srn}).
\end{itemize}

After saving the above parameters in a JSON-formatted configuration file\footnote{An example of a \scope configuration file can be found at \url{https://github.com/wineslab/colosseum-scope/blob/main/radio_api/radio_interactive.conf}.} (e.g., named \texttt{radio.conf}), experiments can be started through the commands shown in Listing~\ref{lst:scope-start}.
\begin{lstlisting}[language=mybash,style=mystyle,
caption={Commands to start \scope applications.},
label={lst:scope-start}]
#!/bin/bash
cd radio_api/
python3 scope_start.py --config-file radio.conf
\end{lstlisting}
This command, executed on the \glspl{srn} assigned to the user, takes care of starting base station, core network and \glspl{ue} applications.

At experiment run time, the \scope \glspl{api} can be used to reconfigure the base station, e.g., to modify the amount of \glspl{rbg} allocated to each slice, or their scheduling policy (see~\cite[Section~3.3]{bonati2021scope}).
Finally, \gls{ran} \glspl{kpm} are automatically logged by the base stations and saved into CSV-formatted files.
These files can either be used on-the-fly (e.g., to perform online inference or model training) or retrieved at a later time from Colosseum data storage (e.g., for offline training or data processing).

\section{O-RAN Control Architecture}
\label{sec:coloran}

The O-RAN control architecture component used by \openrangym is provided by \coloran, an open development environment to design, train and test data-driven O-RAN-compliant solutions at scale~\cite{polese2021coloran}.
\coloran offers a minimalist  implementation of the \gls{osc} near-RT \gls{ric}, that can be instantiated on Colosseum through Docker containers, as well as scripts for the automated deployment of the \gls{ric} components.
\coloran includes containers for the O-RAN \textit{E2 termination, E2 manager} and \textit{E2 routing manager} that handle the communications within the \gls{ric} and with the \gls{ran} nodes (e.g., \scope base stations), a \textit{Redis database} that records the \gls{ran} nodes connected to the \gls{ric}, and an \textit{xApp \gls{sdk}} (Fig.~\ref{fig:openran-architecture}).
The latter provides software tools to design and test \gls{ai}/\gls{ml}-based xApps for run-time \gls{ran} inference and/or control.

\begin{figure}[ht]
    \centering
    \includegraphics[width=\columnwidth]{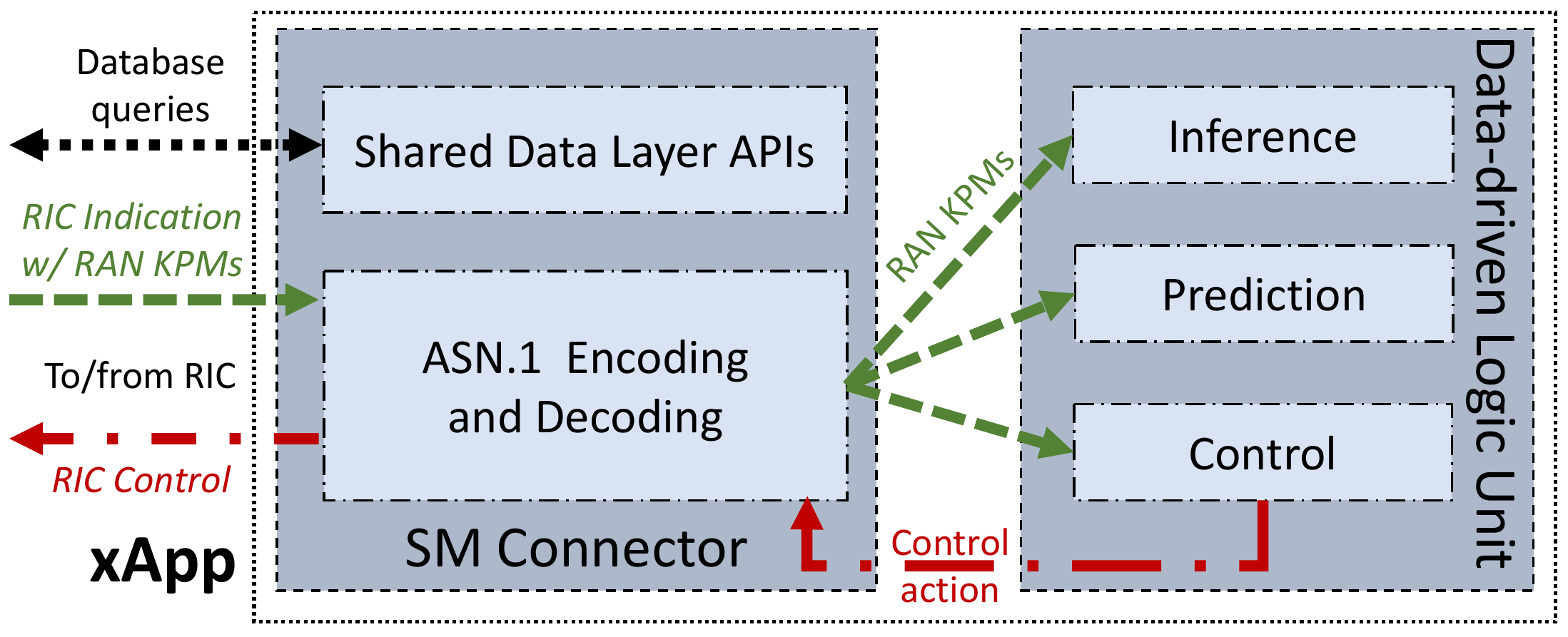}
    \caption{\coloran xApp, adapted from~\cite{polese2021coloran}.}
    \label{fig:xapp}
\end{figure}
At a high level, \coloran xApps are made of two building blocks, shown in Fig.~\ref{fig:xapp}: (i)~the \textit{\gls{sm} connector}, which handles the communications to/from the near-RT \gls{ric} (e.g., to communicate with the base stations), ASN.1 message encoding/decoding, and queries to the \gls{ric} Redis database, and (ii)~the \textit{data-driven logic unit} that performs tasks based on \glspl{kpm} received from the \gls{ran} at run time, e.g., traffic prediction and/or control of the base stations. Notice that at the time of this writing, \coloran xApps use custom \glspl{sm}. Standard-compliant \glspl{sm} are part of our future work.

In the remainder of this section, we will show how to instantiate the \coloran near-RT \gls{ric} on Colosseum (Section~\ref{sec:near-rt-ric-start}), connect the \scope base station to the \gls{ric} through the O-RAN E2 termination (Section~\ref{sec:scope-ric-connect}), and start a sample xApp that interfaces with the base station (Section~\ref{sec:setting-xapp}).
Even though we focus on the implementation for the Colosseum environment (provided in a ready-to-use container image, namely \texttt{coloran-near-rt-ric}), 
\coloran can be ported to different platforms with minor adjustments (see~\cite[Section~6]{polese2021coloran}).

\subsection{Starting the \coloran Near-RT RIC}
\label{sec:near-rt-ric-start}

%
The Docker images of the \coloran near-RT \gls{ric} can be built and started as containers through the provided \texttt{setup-ric.sh} script and the commands of Listing~\ref{lst:ric-start}.
This script (adapted from~\cite{johnson-powder-ric-profile}) takes as argument the network interface used by the \gls{ric} to communicate with the \gls{ran} (e.g., the \texttt{col0} interface in Colosseum).
\begin{lstlisting}[language=mybash,style=mystyle,
caption={Commands to set up the \coloran near-RT \gls{ric}.},
label={lst:ric-start}]
#!/bin/bash
cd setup-scripts/
./setup-ric.sh col0
\end{lstlisting}
First, base Docker images, used to build the \gls{ric} images, are imported.
Then, the four images composing the near-RT \gls{ric} are built, and their IP addresses and ports (defined in the \texttt{setup-lib.sh} script) are configured.
These images include: (i)~\texttt{e2term}, which is the endpoint of the E2 messages on the \gls{ric} (``E2 Termination'' in Fig.~\ref{fig:openran-architecture});
(ii)~\texttt{e2mgr}, which manages the messages to/from the E2 interface (``E2 Manager'');
(iii)~\texttt{e2rtmansim}, which leverages the \gls{rmr} protocol to route the E2 messages inside the \gls{ric} (``E2 Routing Manager''),
and (iv)~\texttt{db}, which maintains a database of the \gls{ran} nodes connected to the \gls{ric} (``Redis Database'').
After the building process completes, the Docker images are initialized as containers on a Colosseum \gls{srn}, and listen for incoming connections from \gls{ran} nodes implementing an E2 termination endpoint.
The container logs can be read through the \texttt{docker logs} command, e.g., \texttt{docker logs e2term -f} shows the logs of the E2 termination (\texttt{e2term}) container.

\subsection{Connecting the \scope Base Station to \coloran}
\label{sec:scope-ric-connect}

After the \coloran neat-RT \gls{ric} is started following the steps of Section~\ref{sec:near-rt-ric-start}, the \scope base station (set up in Section~\ref{sec:scope-setup}) can be connected to it.
To this aim, the \scope base station runs an instance of the O-RAN E2 termination, which we adapted from the \gls{osc} \gls{du} implementation~\cite{osc-du-l2}.
After connecting to the near-RT \gls{ric}, this component can exchange messages with the xApps running therein.
Specifically, the E2 termination of the base station can: (i)~receive \textit{\gls{ric} Subscription} messages from the xApps; (ii)~send periodic \gls{ran} \glspl{kpm} to the xApps through \textit{\gls{ric} Indication} messages; (iii)~receive \textit{\gls{ric} Control} messages from the xApps, and (iv)~interface with the \scope \glspl{api} to modify the scheduling and slicing configurations of the base station based on the received xApp control. 

Listing~\ref{lst:odu-start} shows the steps to initialize the E2 termination on the \scope base station.
\begin{lstlisting}[language=mybash,style=mystyle,
%aboveskip=10pt,
caption={Commands to build and start the \scope E2 termination process.},
label={lst:odu-start}]
#!/bin/bash
cd colosseum-scope-e2/
./build_odu.sh clean
./run_odu.sh
\end{lstlisting}
The E2 termination is first built through the \texttt{build\_odu.sh} script of line~3, which also sets the IP address and port of the near-RT \gls{ric} to connect to, as well as the network interface used for the connection to the \gls{ric}.
Then, the E2 termination can be started through the \texttt{run\_odu.sh} script (line~4), which initializes the E2 termination and connects it to the near-RT \gls{ric}.
The successful connection of base station and near-RT \gls{ric} can be verified by reading the logs of the \texttt{e2term} container (through the command \texttt{docker logs e2term -f}, see Section~\ref{sec:near-rt-ric-start}).
This log shows the association messages between the \gls{ric} and the base station, together with the ID of the connected base stations (e.g., \texttt{gnb:311-048-01000501}).

\begin{figure*}[t]
\setlength\belowcaptionskip{-10pt}
    \centering
    \includegraphics[width=.94\textwidth]{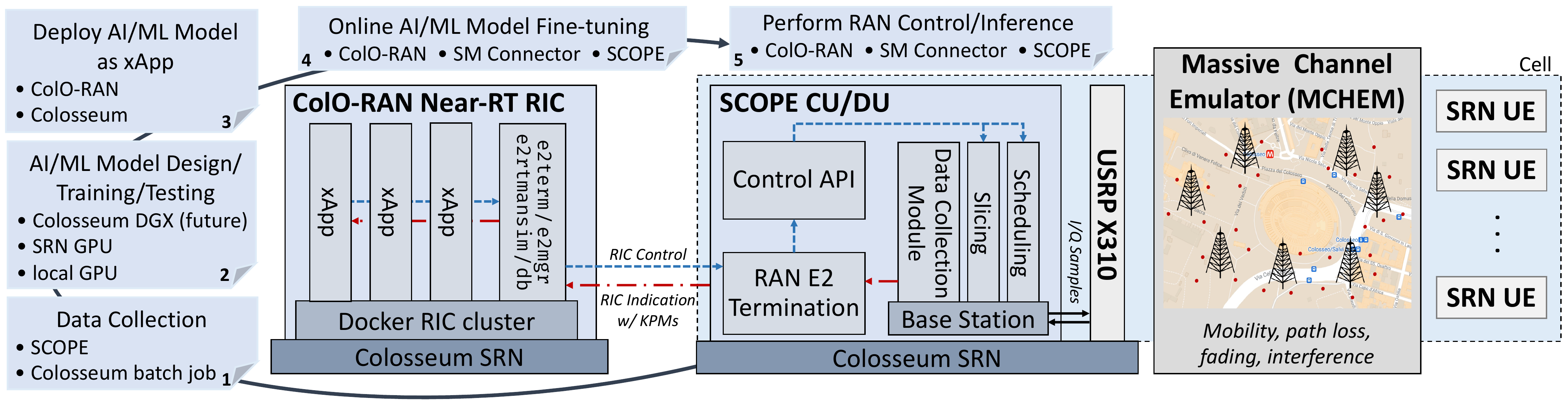}
    \caption{xApp design and testing workflow in \openrangym.}
    \label{fig:xapp-workflow}
\end{figure*}

\subsection{Initializing a Sample xApp}
\label{sec:setting-xapp}

After starting the near-RT \gls{ric}, and connecting the \scope base station to it, the sample xApp provided as part of \coloran can be initialized.
This can be done through the \texttt{setup-sample-xapp.sh} script and the commands shown in Listing~\ref{lst:xapp-build}.
\begin{lstlisting}[language=mybash,style=mystyle,
caption={Commands to build the \coloran sample xApp Docker image, and to start and configure the xApp container.},
label={lst:xapp-build}]
#!/bin/bash
cd setup-scripts/
./setup-sample-xapp.sh gnb:311-048-01000501
\end{lstlisting}
The script takes as input the ID of the \gls{ran} node the xApp subscribes to (e.g., the base station), which can be read in the logs of the \coloran \texttt{e2term} Docker container (see Section~\ref{sec:scope-ric-connect}).
It then builds the Docker image of the sample xApp, and starts it as a Docker container on the near-RT \gls{ric}.

After the xApp container (named, for instance \texttt{sample-xapp}) has been started, the xApp process can be run with the commands shown in Listing~\ref{lst:xapp-run}.
%
\begin{lstlisting}[language=mybash,style=mystyle,
caption={Commands to run the \coloran sample xApp process.},
label={lst:xapp-run}]
#!/bin/bash
docker exec -it sample-xapp /home/sample-xapp/run_xapp.sh
\end{lstlisting}
By running these commands, the xApp subscribes to the \gls{ran} node specified in Listing~\ref{lst:xapp-build} (through a \gls{ric} Subscription message), and triggers periodic reports (sent through \gls{ric} Indication messages) of \gls{ran} \glspl{kpm} from the node.
%

After performing these steps, the \coloran sample xApp logs on file the \glspl{kpm} received from the \gls{ran} node.
Users of \openrangym can add custom intelligence (e.g., through \gls{ai}/\gls{ml} agents) to the xApp by modifying the template scripts in the \texttt{setup/sample-xapp/} directory, and rebuilding the xApp Docker image through the steps described in this section.

\section{xApp Design and Testing Workflow}
\label{sec:xapp-design-testing-workflow}

The steps of the workflow to develop a data-driven xApp in \openrangym, shown in Fig.~\ref{fig:xapp-workflow}, are described next.

1)~\textbf{Data collection.} Data to train the \gls{ai}/\gls{ml} model that will be included in the xApp needs to be collected.
In Colosseum, this can be done by using \scope to instantiate a large-scale cellular network with multiple base stations and \glspl{ue} (see Section~\ref{sec:scope} and~\cite{bonati2021scope}).
%
%
During the experiment, the base stations log the \glspl{kpm} relative to the performance of the served \glspl{ue}, thus creating CSV-formatted datasets representative of the \gls{ran} statistics. Colosseum then transfers these datasets to its internal data storage, making them accessible to the users after the experiment ends.
%
%
Data should be representative of a variety of environments and conditions, so that the
model can adapt to diverse channel conditions and traffic requirements.
Using \openrangym within Colosseum, this can be achieved by running several experiments that emulate different wireless and traffic scenarios (e.g., through \scope cellular scenarios~\cite{bonati2021scope}).

However, manually running hundreds of experiments at scale and in different scenarios is not trivial, and it may take a long time.
To facilitate this task, users can leverage Colosseum \textit{batch jobs}, which take care of automatically running scheduled experiments (or jobs) configured through JSON files~\cite[Section~IV]{bonati2021colosseum}.
In this way, users can schedule experiments in different scenarios and configuration in advance, and then retrieve the generated data from Colosseum data storage.
%
%

2)~\textbf{\gls{ai}/\gls{ml} model design, training and testing.} After collecting data in the desired wireless 
environments, the
model can be designed.
This step involves selecting the \gls{ai}/\gls{ml} techniques that the model will use, which data it should take as input, the reward function of the model, and the set of actions
produced as output (e.g., to perform inference or control of the \gls{ran}).

After the model has been designed, it can be trained and tested offline on the data collected in step~1.\footnote{It is worth mentioning that the O-RAN specifications do not permit the deployment of \gls{ai}/\gls{ml} models that have not been pre-trained offline. This is to shield the \gls{ran} from poor performance or outages~\cite{oran-wg2-ml}.}
%
These phases, which usually benefit from \gls{gpu}-enabled environments, can either be carried out locally (on the user's own \glspl{gpu}), or on the \glspl{gpu} of the \glspl{srn}.
As another option to train and test \gls{ai}/\gls{ml} solutions, Colosseum has recently added two NVIDIA DGX servers with A100 \glspl{gpu} as part of its planned extensions~\cite{bonati2021colosseum}.
Although at the time of this writing these resources cannot be reserved by the users of the testbed yet, their general availability is envisioned for Q2~2022.
These novel servers will significantly increase Colosseum's computational capabilities, making it, together with \openrangym,  a key tool to develop data-driven solutions to control Open RAN systems.

3)~\textbf{Deploy the \gls{ai}/\gls{ml} model as an xApp.} After the
model has been designed, trained, and tested,
it can be deployed as an xApp on the \coloran near-RT \gls{ric}.
This can be done through the procedures described in Section~\ref{sec:setting-xapp}.
Specifically, the trained \gls{ai}/\gls{ml} model can be included in the \coloran sample xApp (as the \textit{data-driven logic unit} of Fig.~\ref{fig:xapp}) modifying the template scripts in the \texttt{setup/sample-xapp/} directory.
Then, the Docker image of the xApp with the user-defined logic is built with the commands of Listing~\ref{lst:xapp-build}, and instantiated on the \coloran near-RT \gls{ric} through the commands of Listing~\ref{lst:xapp-run}.

4)~\textbf{Online \gls{ai}/\gls{ml} model fine-tuning.} Upon startup, the xApp interfaces with the \scope base station through the \coloran near-RT \gls{ric} and the O-RAN E2 termination.
First, the xApp subscribes to \scope \gls{cu}/\gls{du} by sending it \gls{ric} Subscription messages.
Then, it triggers periodic reports (tunable based on the experiment requirements~\cite{polese2021coloran}) of \gls{ran} \glspl{kpm} from the base station, sent through \gls{ric} Indication messages.
The xApp may also use the \glspl{kpm} in these report messages to fine-tune the model online, which allows it to adapt to the actual production environment where it will be deployed. Once the model has undergone this additional phase of online training, the xApp Docker image can be updated to save the newly trained
weights of the model.

5)~\textbf{Perform \gls{ran} control/inference.} Now the xApp can be used in the production infrastructure to perform run-time inference and control of the \gls{ran}.
This latter step is achieved by having the xApp transmitting the actions computed by the \gls{ai}/\gls{ml} model (e.g., to modify the parameters and configuration of the base station) through \gls{ric} Control messages.
These are sent to the base station through the O-RAN E2 interface, where they are processed by the \gls{cu}/\gls{du} the xApp is subscribed to.
At the \gls{cu}/\gls{du} side, these messages trigger the \scope control \glspl{api} (see Fig.~\ref{fig:xapp-workflow}) that apply the newly received configuration to the protocol stack of the base station at run time.

\smallskip
We now showcase an example of xApps designed and tested with \openrangym, and used to control a cellular network with 7~base stations and 42~\glspl{ue} (6~\glspl{ue} per base station) instantiated on Colosseum.
Each base station serves \glspl{ue} with different \gls{qos} requirements, divided on three network slices: \gls{embb}, \gls{mtc}, and \gls{urllc} slice.
We follow the workflow described in this section to design two \gls{drl}-based xApps---trained on a collected dataset with $3.4$\:GB of \gls{ran} traces, and more than 73~hours of experiments---that control the configuration of the base stations at run time using \gls{ran} \glspl{kpm}, as discussed in~\cite{polese2021coloran}.
%
One xApp (named \texttt{sched}) manages the scheduling policies of the base station; the other (\texttt{sched-slicing}) also allocates the amount of resources available to each slice.

\begin{figure}[ht]
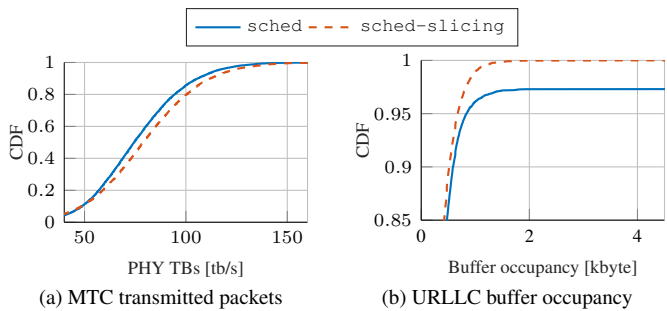

\setlength\abovecaptionskip{0.1cm}
\setlength\belowcaptionskip{-1pt}
    \centering
    \ifexttikz
        \tikzsetnextfilename{sched-vs-sched-slicing-pkt-slice-1}
    \fi
    \begin{subfigure}[b]{0.48\columnwidth}
        \centering
        \setlength\fwidth{.8\columnwidth}
        \setlength\fheight{.5\columnwidth}
        \input{figures/sched-vs-sched-slicing-pkt-slice-1.tex}
       \setlength\abovecaptionskip{-0.4cm}
        \caption{MTC transmitted packets}
        \label{fig:xapp-pkts-mtc}
    \end{subfigure}\hfill
    \ifexttikz
        \tikzsetnextfilename{sched-vs-sched-slicing-buffer-slice-2}
    \fi
    \begin{subfigure}[b]{0.48\columnwidth}
        \centering
        \setlength\fwidth{.8\columnwidth}
        \setlength\fheight{.5\columnwidth}
        \input{figures/sched-vs-sched-slicing-buffer-slice-2.tex}
       \setlength\abovecaptionskip{-0.4cm}
        \caption{URLLC buffer occupancy}
        \label{fig:xapp-buffer-urllc}
    \end{subfigure}\hfill
    \caption{Comparison of xApps developed with \openrangym.}
    \label{fig:xapp-comparison}
\end{figure}

Figure~\ref{fig:xapp-comparison} illustrates the \gls{cdf} of the \gls{ran} when the two xApps are instantiated on the \coloran near-RT \gls{ric} and used to control the network.
Statistics on the transmitted \glspl{tb} for the \gls{mtc} slice are shown in Fig.~\ref{fig:xapp-pkts-mtc}; on the downlink buffer occupancy of the \gls{urllc} slice in Fig.~\ref{fig:xapp-buffer-urllc}.
In this example, both xApps aim at maximizing the transmit rate of the \gls{mtc} traffic, and at minimizing the time packets remain in the base station queues for the \gls{urllc} traffic.
%
By acting on an additional action set (i.e., the slice resource allocation), the \texttt{sched-slicing} xApp achieves better performance both in terms of transmitted packets and of buffer occupancy.\footnote{A detailed evaluation of \openrangym xApps, including their orchestration, and control of large-scale experimental networks can be found in~\cite{polese2021coloran, doro2022orchestran}.}

%
%

\section{Conclusions}
\label{sec:conclusions}

We presented \openrangym, the first publicly-available research platform for data-driven O-RAN experimentation at scale.
Building on frameworks for data collection and \gls{ran} control, \openrangym enables end-to-end design and testing of data-driven O-RAN xApps.
We described the main components of \openrangym, detailing configuration options and procedures for experimenting at scale on Colosseum.
We then gave an overview of the \openrangym xApp design and testing workflow, from user instantiation of their \gls{ai}/\gls{ml} models as xApps on the near-RT \gls{ric}, to performing inference, prediction and/or control of base stations using live \glspl{kpm} from the \gls{ran}.
Finally, we provided an example of two xApps designed with \openrangym and used to control a large-scale O-RAN-managed network deployed on Colosseum.
\openrangym has been made publicly-available to the research community, and opened up for community contributions and additions.

\footnotesize  
\bibliographystyle{IEEEtran}
\bibliography{biblio.bib}

\end{document}